# Speculative Automated Refactoring of Imperative Deep Learning Programs to Graph Execution

Raffi Khatchadourian,*† Tatiana Castro Vélez,† Mehdi Bagherzadeh,‡ Nan Jia,† Anita Raja*†
*City University of New York (CUNY) Hunter College, †CUNY Graduate Center, ‡Oakland University
Email: khatchad@hunter.cuny.edu, tcastrovelez@gradcenter.cuny.edu, mbagherzadeh@oakland.edu,
njia@gradcenter.cuny.edu, anita.raja@hunter.cuny.edu

***Abstract*—Efficiency is essential to support ever-growing datasets, especially for Deep Learning (DL) systems. DL frameworks have traditionally embraced *deferred* execution-style DL code—supporting symbolic, graph-based Deep Neural Network (DNN) computation. While scalable, such development is error-prone, non-intuitive, and difficult to debug. Consequently, more natural, imperative DL frameworks encouraging *eager* execution have emerged but at the expense of run-time performance. Though hybrid approaches aim for the "best of both worlds," using them effectively requires subtle considerations. Our key insight is that, while DL programs typically execute sequentially, hybridizing imperative DL code resembles parallelizing sequential code in traditional systems. Inspired by this, we present an automated refactoring approach that assists developers in determining which otherwise eagerly-executed imperative DL functions could be effectively and efficiently executed as graphs. The approach features novel static imperative tensor and side-effect analyses for Python. Due to its inherent dynamism, analyzing Python may be unsound; however, the conservative approach leverages a *speculative* (keyword-based) analysis for resolving difficult cases that informs developers of any assumptions made. The approach is: (i) implemented as a plug-in to the *PyDev Eclipse* IDE that integrates the *WALA Ariadne* analysis framework and (ii) evaluated on nineteen DL projects consisting of 132 KLOC. The results show that 326 of 766 candidate functions (42.56%) were refactorable, and an average relative speedup of 2.16x on performance tests was observed with negligible differences in model accuracy. The results indicate that the approach is useful in optimizing imperative DL code to its full potential.**

***Index Terms*—deep learning, refactoring, imperative programs, hybrid programming paradigms, graph execution**

## I. Introduction

Machine Learning (ML), including Deep Learning (DL), systems are pervasive. They use dynamic models whose behavior is ultimately defined by input data; however, as datasets grow, efficiency becomes essential [5]. DL frameworks have traditionally embraced a *deferred* execution-style that supports symbolic, graph-based Deep Neural Network (DNN) computation [6], [7]. While scalable, such development is error-prone, cumbersome, and produces programs that are difficult to debug [8]–[11]. Contrarily, more natural, less error-prone, and easier-to-debug *imperative* DL frameworks [12]–[14] encouraging *eager* execution have emerged. Though ubiquitous, such programs are less efficient and scalable than their deferred-execution counterparts [7], [13], [15]–[18]. Thus, hybrid approaches [15], [16], [19] execute imperative DL programs as static graphs at run-time. For example, in *TensorFlow* [20], *AutoGraph* [15] can enhance performance[1] by decorating (annotating) appropriate Python functions with `@tf.function` (q.v. §III) [21]. Developers can use `tf.function` to make graphs out of their programs; it is a transformation tool that creates Python-independent dataflow graphs out of their Python code, helping developers create performant yet portable models [22]. Consequentially, decorating functions with such hybridization APIs can increase (otherwise eagerly-executed) imperative DL code performance without explicit code modification.

Though promising, hybridization necessitates non-trivial metadata [17] and exhibits limitations and known issues with native program constructs [22]; using them effectively requires subtle considerations [23]–[25]. Khatchadourian et al. [26], [27] only briefly present preliminary progress and engineering aspects, respectively, towards this end. Alternative approaches [17], [28], [29] may impose custom Python interpreters or require additional or concurrently running components, which may be impractical for industry, support only specific Python constructs, or still require function decoration. Thus, developers are burdened with *manually* specifying the functions to be converted. Manual analysis and refactoring can be overwhelming, prone to errors and omissions [30], and complicated by Object-Orientation (OO) [14].

We propose an automated refactoring approach that transforms otherwise eagerly-executed imperative (Python) DL code for enhanced performance by specifying whether such code could be effectively and efficiently executed as graphs at run-time. Our key insight is that, while imperative DL programs typically execute sequentially, hybridizing such code resembles parallelizing sequential code in traditional software. For example, to avoid unexpected behavior, like concurrent programs, hybrid functions should avoid side-effects.

Inspired by such refactorings [31], our approach is based on a novel tensor (matrix-like data structures) analysis for imperative DL code and a thorough investigation of the *TensorFlow* documentation [22]. The approach infers when eagerly-executed imperative DL code can effectively and efficiently execute as graphs. This determination depends on which of

This material is based upon work supported by the National Science Foundation under Award Nos. CCF-22-00343, CNS-22-13763, and CCF-23-43750.

[1] "Performance" in this paper refers to *run-time* performance (speed), not model accuracy.

our defined refactoring preconditions pass. The approach also determines when graph execution may be counterproductive.

The approach features a side-effect analysis for Python. Due to the inherent dynamism of Python, our analysis may be unsound; however, our conservative approach adopts a *speculative*, keyword-based analysis [5], [32] that incorporates domain knowledge to resolve difficult cases. Developers are informed of any assumptions made during the speculative analysis, allowing them to validate these assumptions if necessary (q.v. § IV-A and IV-C2). While the refactorings operate on imperative DL code that is easier-to-debug than its deferred-execution counterparts, the refactorings themselves do not improve debuggability. Instead, they facilitate *performant* easily-debuggable (imperative) DL code.

While Large Language Models (LLMs) [33] and big data-driven refactorings [34] have emerged, obtaining a sufficiently large and correct dataset to automatically extract the proposed refactorings is challenging. This difficulty arises because developers struggle with manually migrating DL code to graph execution [23]. Moreover, due to our enhancements to *Ariadne* [4]—a tool for static analysis of tensor shapes originally designed for deferred execution-style Python DL code—our interprocedural analysis works with *complete* projects spanning multiple files and directories. It is not bound by prompt token size restrictions [35]. Although developers generally underuse automated refactorings [36], [37], data scientists and engineers may not be classically trained software engineers. Therefore, they may be more open to using automated (refactoring) tools to develop software [38]–[40]. Furthermore, our approach is fully automated with minimal barrier to entry.

Our refactoring approach is implemented as an open-source *PyDev* [1] *Eclipse* [2] Integrated Development Environment (IDE) plug-in [27] that integrates static analyses from *WALA* [3]—a library of many common program analyses—and *Ariadne*. We build atop *Ariadne* as it is a widely-used [41]–[43] static analysis framework that supports Python and tensor analysis, which is a linchpin for our approach. It translates its intermediate representation (IR) to *WALA*, which supports many static analyses. These include call graph (CG) construction and ModRef analysis used for side-effect analysis (q.v. §IV-E). *Ariadne* also supports several popular dynamic features, e.g., function callbacks (q.v. §IV-C3b). While alternatives [41], [44] exist, they are either also built atop *Ariadne* or do not support the same level of tensor analysis at this time. And, since *Ariadne* is written in Java, it is easier to integrate with *PyDev*. While dynamic analyses [45], [46] exist, static analysis can cover the large combinatorial space imposed by the numerous parameters and possible inputs to a DNN [47]. Furthermore, identifying run-time performance bottlenecks often requires executing the entire program—a potentially lengthy process due to training [24]. Lastly, refactorings (source-to-source program transformations) must work on some level of static information, as the source code is ultimately transformed.

To evaluate the approach, we applied our plug-in on nineteen Python imperative DL programs of varying size and

```
1 # Build a graph.
2 a = tf.constant(5.0)
3 b = tf.constant(6.0)
4 c = a * b
5 # Launch graph in a session.
6 sess = tf.Session()
7 # Evaluate the tensor `c`.
8 print(sess.run(c)) # prints 30.0
```

Listing 1: *TensorFlow* deferred execution-style code.

domain with a total of ∼132 thousand lines of code. Due to its popularity and extensive use by previous work [8], [9], [11], [48]–[53], we focus on hybridization in *TensorFlow* but also discuss generalization in §IV-G. Our study indicates that: (i) the interprocedural, fully automated analysis cost is reasonable, with an average running time of 0.17 s per candidate function and 11.86 s per thousand lines of code, (ii) despite its ease-of-use, `tf.function` is not commonly (manually) used in imperative DL software, motivating an automated approach, and (iii) the proposed approach is useful in refactoring imperative DL code for greater efficiency despite being conservative. This work makes the following contributions:

**Precondition formulation.** We present a novel refactoring approach for maximizing the efficiency of imperative DL code by automatically determining when such functions can execute as graphs and when graph execution may be counterproductive. Our approach refactors imperative DL code for enhanced performance with negligible differences in model accuracy.

**Modernization of *Ariadne* for imperative DL.** We modernize *Ariadne* by adding static analyses of tensors in imperative DL programs, new Python language features, Python side-effects, and additional modeling.

**Implementation and experimental evaluation.** To ensure real-world applicability, the approach was implemented as *PyDev Eclipse* IDE plug-in built on *WALA* and *Ariadne* and used to study nineteen Python DL programs. It successfully refactored 42.56% of candidate functions, and we observed an average relative speedup of 2.16x during performance testing. The experimentation also sheds light onto how hybridization is used, potentially motivating future language and API design. The (publicly available [54]) results advance the state-of-the-art in automated tool support for imperative DL code to perform to its full potential.

## II. Background

*Deferred* execution-style APIs make DNNs straight-forward to execute as symbolic graphs that enable run-time optimizations. For example, during graph building (lines 2–4 of Listing 1), line 4 does not execute until the `Session` created on line 6 is run on line 8. While efficient, legacy code using such APIs is cumbersome, error-prone, and difficult to debug and maintain [8]–[11]. Such APIs also do not natively support common imperative program constructs, e.g., iteration. Contrarily, *eager* execution-style APIs [12], [13] facilitate imperative and OO [14] DL programs that are easier-to-debug, less error-prone, and more extensible. For instance, under eager execution, line 4 of Listing 1 would execute *immediately* to evaluate the tensor `c`.

```
1
2 class SequentialModel(Model):
3   def __init__(self, **kwargs):
4     super(SequentialModel, self)
5       .__init__(...)
6     self.flatten = layers.Flatten(
7       input_shape=(28, 28))
8     num_layers = 100 # Add layers.
9     self.layers = [layers
10      .Dense(64,activation="relu")
11       for n in range(num_layers)]
12    self.dropout = Dropout(0.2)
13    self.dense_2 = layers.Dense(10)
14
15
16  def __call__(self, x):
17    x = self.flatten(x)
18    for layer in self.layers:
19      x = layer(x)
20    x = self.dropout(x)
21    x = self.dense_2(x)
22    return x
23
24 d = tf.random.uniform([20,28,28])
25 model = SequentialModel()
26 model(d)
```

(a) Code snippet before refactoring.

```
1 import tensorflow as tf
2 class SequentialModel(Model):
3   def __init__(self, **kwargs):
4     super(SequentialModel, self)
5       .__init__(...)
6     self.flatten = layers.Flatten(
7       input_shape=(28, 28))
8     num_layers = 100 # Add layers.
9     self.layers = [layers
10      .Dense(64,activation="relu")
11       for n in range(num_layers)]
12    self.dropout = Dropout(0.2)
13    self.dense_2 = layers.Dense(10)
14
15  @tf.function
16  def __call__(self, x):
17    x = self.flatten(x)
18    for layer in self.layers:
19      x = layer(x)
20    x = self.dropout(x)
21    x = self.dense_2(x)
22    return x
23
24 d = tf.random.uniform([20,28,28])
25 model = SequentialModel()
26 model(d)
```

(b) Improved code via refactoring.

Listing 2: *TensorFlow* imperative (OO) DL model code [18].

```
1 def f(x):
2   print(x)
3 f(1)
4 f(1)
5 f(2)
```

(a) Code before refactoring.

```
1
1
2
```

(b) Output before.

```
1
2
```

(c) Hypothetical output.

```
1 def f(x):
2   print(x)
3 f(1)
4 f(1)
5 f(2)
```

(d) Safely unrefactored code.

Listing 3: Imperative *TensorFlow* code with Python side-effects [22].

## III. Motivation

Listing 2a portrays *TensorFlow* imperative (OO) DL code representing a modestly-sized model for classifying images. By default, it runs eagerly; however, it may be possible to enhance performance by executing it as a graph at run-time. Listing 2b, lines 1 and 15, show the refactoring with the imperative DL code executed as a graph at run-time (added code is underlined). *AutoGraph* [15] is now used to potentially improve performance by decorating `__call__()`—a special function that converts objects to functors, i.e., "callables," e.g., line 26—with `@tf.function`. At run-time, `__call__()`'s execution will be "traced, " producing an equivalent graph [22]. A relative speedup of ∼9.22 ensues [21].

Though promising, using hybridization effectively and efficiently is challenging [17], [22], [23]. Developers face several key challenges when manually hybridizing DL code: understanding when hybridization is beneficial, identifying which functions are safe to hybridize, and avoiding semantic changes that break program correctness. These challenges stem from the complex interactions between the Python language and graph execution constraints.

For instance, side-effect producing, native Python statements, e.g., printing, list appending, global variable mutation, are problematic for `tf.function`-decorated functions, i.e., "`tf.functions`" [22]. Because their executions are traced, a function's behavior is "etched" (frozen) into its corresponding graph and thus may exhibit unexpected results. Side-effects may be executed multiple times or not at all. Side-effects occur when `tf.functions` are called the first time; subsequent calls with similar arguments execute the graph instead. For example,

```
1 class Model(tf.Module):
2   def __init__(self):
3     self.v=tf.Variable(0)
4     self.counter = 0
5
6   def __call__(self):
7     if self.counter == 0:
8       self.counter += 1
9       self.v.assign_add(1)
10    return self.v
11
12 m = Model()
13 for n in range(3):
14   print(m().numpy())
```

(a) Code snippet before refactoring.

```
1
1
1
```

(b) Output before refactoring.

```
1
2
3
```

(c) Hypothetical output.

```
1 class Model(tf.Module):
2   def __init__(self):
3     self.v=tf.Variable(0)
4     self.counter = 0
5
6   def __call__(self):
7     if self.counter == 0:
8       self.counter += 1
9       self.v.assign_add(1)
10    return self.v
11
12 m = Model()
13 for n in range(3):
14   print(m().numpy())
```

(d) Safely unrefactored code.

Listing 4: Imperative *TensorFlow* code using a counter [22].

on line 2 of Listing 3a, `f()` outputs `x`. Then, `f()` is invoked three times, the first two with the argument `1` and the last with `2`. The corresponding output is shown in Listing 3b. Note that this code executes eagerly.

Unlike the previous example, migrating this code to a graph at run-time—by decorating `f()` with `@tf.function`—could be counterproductive because it would alter the original program semantics. If we hybridize `f()`, the output would instead be that shown in Listing 3c. The reason is that the first invocation of `f()` on line 3 would result in a graph being built (through tracing) that—due to a similar argument—is later used on line 4. Consequently, the side-effecting code on line 2 would *not* be exercised. In contrast, line 2 *is* exercised as a result of the call on line 5 due to a different argument. As such, the code in Listing 3d remains eagerly-executed as semantics must be preserved.

Although Listing 3 is simple, avoiding unexpected behavior caused by refactoring can be difficult. Consider Listing 4a, where a model uses a `counter` to safeguard a variable incrementation, and its corresponding output in Listing 4b. Like Listing 2a, the model's `__call__()` method is executed eagerly. Unlike Listing 2b, however, decorating `__call__()` with `@tf.function` would alter semantics—the output would be that shown in Listing 4c. The reason is that the initial value of `counter` is captured during tracing upon the first model invocation (line 14 of Listing 4a). The overall effect is that the value of `v` is incremented *unconditionally* (line 9) each time the model is invoked. Thus, the code should remain unrefactored, as depicted in Listing 4d. Such problems are common when (manually) migrating imperative DL code to graph execution [23]. Worse yet, developers only realize such errors after refactoring and subsequently observing suspicious numerical results or significantly lower performance than expected (e.g., when guarded operations are costly) [22].

Besides ensuring that DL code is amenable to hybridization [55], developers must also know *when* and *where* to use it to avoid performance bottlenecks and other undesired behavior. For example, confusion exists on how often `@tf.function` should be applied [56], and calling `tf.functions` recursively could cause infinite loops [22]. Even if a recursion seems to work, the `tf.function` will be traced *multiple* times ("retracing"), potentially impacting performance. Using `tf.function` on small computations can be dominated by graph creation overhead [18]. Thus, care should be taken to

```python
1 @tf.function
2 def train(num_steps):
3   for _ in tf.range(num_steps):
4     train_one_step()
5
6 train(10)
7 train(20)
8 train(tf.constant(10))
9 train(tf.constant(20))
```

(a) Code snippet before refactoring.

```python
1 @tf.function
2 def train(num_steps):
3   for _ in tf.range(num_steps):
4     train_one_step()
5
6 train(10)
7 train(20)
8 train(tf.constant(10))
9 train(tf.constant(20))
```

(b) Improved code via refactoring.

Listing 5: Imperative *TensorFlow* code using primitive literals [22].

not use hybridization unnecessarily, e.g., on functions that do not receive appropriate (tensor) arguments. Such functions may incur tracing overhead without performance benefits. Developers have struggled with these considerations [24], [25] and have manually analyzed and corrected their code [23].

Retracing helps ensure that the correct graphs are generated for each set of inputs; however, excessive retracing may cause code to run more slowly had `tf.function` *not* been used [22], [57], [58]. Consider Listing 5a that depicts imperative *TensorFlow* code that uses both Python primitive literals (lines 6 and 7) and tensors (lines 8 and 9) as arguments to the `num_steps` parameter of `train()`. On both lines 6 and 8, a new graph is created. However, *another* graph is created on line 7, resulting in a retrace, while the graph created on line 8 is *reused* on line 9. This is due to the rules of tracing [22]; graphs are generated for tensor arguments based on their data type and shape, while for Python primitive values, the scheme is based on the value itself. For example, the `TraceType`—a *TensorFlow* data structure used to determine whether traces can be reused—of the value `3` is `LiteralTraceType<3>` and not `int` [22]. Listing 5b depicts the refactored version (removed code is ~~struck through~~), where `train()` has been de-hybridized (line 1). Note that it is safe to do so as, in contrast to Listing 3, Listing 5 contains no Python side-effects.

These examples demonstrate that effectively using hybridization is not always straight-forward, requiring complex analysis and a thorough API understanding—a compounding problem in more extensive programs. As imperative DL programming becomes increasingly pervasive, it would be extremely valuable to developers/data scientists—particularly those not classically trained software engineers—if automation can assist in writing reliable and efficient imperative DL code.

## IV. Optimization Approach

### A. Hybridization Refactorings

We propose two new refactorings, namely, Convert Eager Function to Hybrid and Optimize Hybrid Function, that—based on a conservative static analysis—transform eager Python functions to hybrid and vice-versa, respectively, when it is potentially advantageous to do so. Our key insight is that, while DL code typically executes sequentially, hybridization shares commonality with concurrent programs. For example, to avoid unexpected behavior, such functions should avoid side-effects (q.v. Listing 3). To avoid (excessive) retracing (q.v. Listing 5), we involve (imperative) tensor analysis to approximate whether functions have tensor parameters. Otherwise, functions may be traced *each* time they are invoked, potentially *degrading* performance [22], [23]. The precondition formulation was inspired by parallelization refactorings of traditional systems [31]. It also involved a thorough study of the *TensorFlow* documentation [22].

TABLE I: Convert Eager Function to Hybrid preconditions.

| | exe | tens | lit* | se | rec | trans |
|---|---|---|---|---|---|---|
| P1 | eag | *T* | *F* | *F* | *F* | hyb |

* An option exists in our implementation to not consider Boolean literals.

Due to the inherent dynamism of Python, our static analysis may be unsound. However, we argue that it is still practically useful as: (i) it is conservative, (ii) advanced language features are not generally extensively used in Python programs [5], [59], (iii) an extensible speculative analysis [5], [32] is used to resolve difficult cases, and (iv) any assumptions made are reported to developers for investigation. While automated analysis of Python is challenging, manual analysis is also difficult. DL systems encompass many functions, files, packages, and dependencies, illuminating a need for automation.

*1) Converting Eager Functions to Hybrid:* Table I depicts the preconditions for the Convert Eager Function to Hybrid refactoring. It lists the conditions that must hold for the transformation to potentially improve run-time performance while averting unexpected behavior. Column **exe** is the current execution mode of the function, either eager (eag) or graph, as determined by the original decorator. Column **tens** is *true* iff the function likely includes "`Tensor`-like" (e.g., `tf.Tensor`, `tf.Variable`) parameters (q.v. §IV-C). Column **lit** is *true* iff the function likely includes a literal passed as an argument to a parameter (q.v. §IV-D). Column **se** is *true* iff the function has Python side-effects (q.v. §IV-E). Column **rec** is *true* iff the function (transitively) recursive (q.v. §IV-F). As mentioned in §III, hybridizing recursive functions may cause infinite loops [22]. Column **trans** is the refactoring action to employ when the corresponding precondition passes (conditions are mutually exclusive).

A function passing P1 is one that is originally executed eagerly, has a tensor argument, does not have a literal (or container—lists, tuples—of literals) argument, has no Python side-effects, and is not recursive. The method defined starting on line 16 of Listing 2a passes P1. Here, there is at least one argument—corresponding to parameter `x`—with type `tf.Tensor` due to the dataflow stemming from the call to `tf.random.uniform()` on line 24. There is also no calls to `__call__()` where `x` takes on a literal argument, e.g., `int`, `float`; such an argument may induce retracing and thus reduce run-time performance (q.v. Listing 5). Furthermore, `__call__()` is not already hybrid, and it does not contain Python side-effects; although it writes to the `x` parameter, `x` here refers to a local copy of the reference variable and later returns its result on line 22. On the other hand, those in Listings 3a and 4a do contain Python side-effects and thus are not refactored as they do not pass P1.

For column **lit**, a common pattern in `Model.call()` func-

TABLE II: OPTIMIZE HYBRID FUNCTION preconditions.

| | exe | tens | lit* | se | trans |
|---|---|---|---|---|---|
| P2 | graph | *F* | N/A | *F* | eag |
| P3 | graph | *T* | *T* | *F* | eag |

* An option exists in our implementation to not consider Boolean literals.

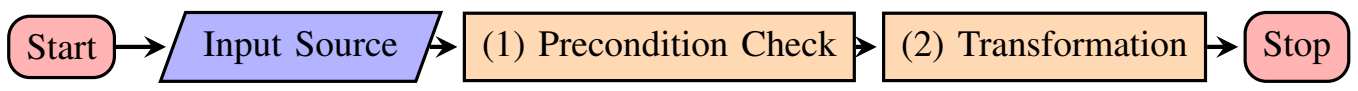

Fig. 2: High-level flowchart.

tions is to pass a Boolean indicating whether the function is called for model training as opposed to model validation. For example, line 6 of Listing 6 applies softmax only when the model is *not* training. Because Booleans can only take on two values, their affect on retracing may be negligible compared to that of other types, e.g., integers (cf. Listing 5). Our implementation (q.v. §V-A) has an option to not consider Booleans during literal inference that we subsequently enable in the evaluation (q.v. §V-B).

*2) Optimizing Hybrid Functions:* Misuses of `tf.function` result in low efficiency [23], [25]. Table II depicts the preconditions for the OPTIMIZE HYBRID FUNCTION refactoring. A function passing P2 is currently executed as a graph but does *not* have a tensor parameter nor contain side-effects. Functions without tensor parameters may not benefit from hybridization—tensor arguments with similar types and shapes potentially enable traces to be reused. P2 also involves checking side-effects. As shown in Listings 3 and 4, hybrid functions with side-effects may produce unexpected results. While converting the function to execute eagerly could potentially stabilize any misbehavior, doing so would not preserve original program semantics. Thus, such functions fail P2, and we warn developers of the situation.

```
1 class NeuralNet(Model): # ...
2   def call(self, x, train=False):
3     x = self.fc1(x)
4     x = self.fc2(x)
5     x = self.out(x)
6     if not is_training:
7       x = tf.nn.softmax(x)
8     return x
```

Listing 6: An NN using Booleans [60].

Note that whether the function is recursive is irrelevant in Table II; if the function has no tensor parameters, de-hybridizing it does not alter semantics as potential retracing happens regardless. However, we warn when a hybrid function with a tensor parameter is recursive. Since hybrid functions passing P2 will be transformed to execute eagerly, it is inconsequential whether it has a literal parameter; retracing occurs only for hybrid functions. P3 de-hybridizes functions to avoid unnecessary retracing, which may cause worse performance had `tf.function` not been used originally [23].

## B. Overview

Figure 2 depicts the high-level flowchart for our approach. The process begins with input source code. Preconditions are checked on the constituent Python function definitions (Step 1). Functions passing preconditions are then transformed to either hybrid or eager by either adding or removing the `@tf.function` function decorator (Step 2).

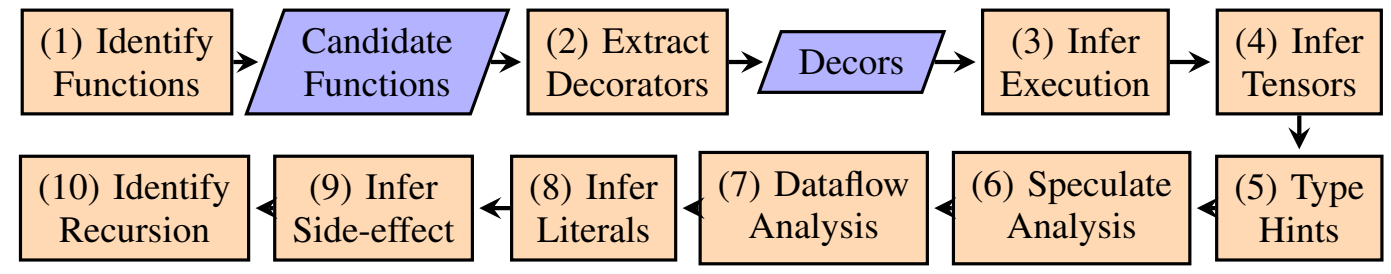

Fig. 3: Precondition checking flowchart.

Precondition checking is further expanded in Fig. 3. First, function definitions are identified (Step 1), producing candidate functions for transformation. Next, function attributes are analyzed, initially by extracting and subsequently examining their function decorators (Step 2). This determines the original function execution mode (Step 3). Tensor parameters are inferred next (Step 4), which includes utilizing Python 3 type hints (Step 5), context-aware speculative analysis (Step 6), and dataflow analysis (Step 7). Literal parameters are inferred next (Step 8), followed by side-effect analysis (Step 9). Finally, recursion is identified (Step 10).

## C. Inferring Tensor Parameters

Hybridization hinges on whether a function likely has parameters of type `Tensor`, a `Tensor`-like type, or a subtype of `Tensor`, e.g., `tf.sparse.SparseTensor`, `tf.RaggedTensor`. We also consider Python *containers* of `Tensor`-like objects. We exclude the implicit parameter `self` as method receivers in this context typically refer to model objects rather than client-side tensors. Three strategies are employed to infer tensor parameters: static (dataflow) analysis, type hints, and speculative analysis.

*1) Type Hints:* Though they are not natively enforced at run-time, if available, we optionally leverage type hints in inferring tensor parameters. If a type hint resolves to a tensor type, we treat the corresponding parameter as a tensor parameter. We also consider containers of `Tensor`-like objects when incorporating type hints.

Type hints may be particularly useful when refactoring library code where client code may not be available; parameter types cannot be inferred via dataflow analysis if there are no calls to the function. Although *TensorFlow* only uses type hints when a specific flag (`experimental_follow_type_hints`) to `tf.function` is provided, we nevertheless provide an option in our implementation to follow type hints regardless of any hybridization arguments (q.v. §V-A).

*2) Speculative Analysis:* We optionally consider function *context*, i.e., speculative analysis [5], [32], when determining likely tensor parameters. This keyword-based approach is only used when: (i) static (dataflow) analysis fails to determine a tensor type, (ii) type hints are unavailable, and (iii) the function has at least one parameter. Previously, the scheme of Zhou et al. [5] was used for procedural, deferred-execution–style DL (*TensorFlow* 1) code (e.g., Listing 1); we adapt it here for *imperative* and OO DL (*TensorFlow* 2) code to inferring tensor parameters. It is mainly used here to prevent de-hybridizing otherwise promising hybrid functions when the

TABLE III: Example tensor "generators."

| API | alias | description | tensl? |
|---|---|---|---|
| `tf.Tensor` | `tf.experimental.numpy.ndarray` | A multidimensional element array. | *F* |
| `tf.sparse.SparseTensor` | `tf.SparseTensor` | A sparse tensor. | *F* |
| `tf.ones` | | Tensor with all ones. | *F* |
| `tf.fill` | | Tensor with a scalar. | *F* |
| `tf.zero` | | Tensor with all zeros. | *F* |
| `tf.one_hot` | | A one-hot tensor. | *F* |
| `tf.eye` | `tf.linalg.eye` | Identity matrix(ces). | *F* |
| `tf.Variable` | | Shared mutable state. | *T* |
| `tf.constant` | | A constant tensor. | *F* |
| `tf.convert_to_tensor` | | Converts a value to `Tensor`. | *F* |
| `tf.keras.Input` | `tf.keras.layers.Input` | Tensor for Keras. | *T* |
| `tf.range`* | | Number sequence. | *F* |

* Generates a tensor containing a *sequence* of tensors.

TABLE IV: Example tensor dataset "generators" (static methods).

| API | description |
|---|---|
| `tf.Dataset.from_tensor_slices` | A dataset from tensor slices. |
| `tf.Dataset.range` | A dataset of a step-separated range of values. |

static analysis cannot infer tensor types. For example, a hybrid function whose name is `training_step` likely deals with tensors. We reuse keywords from Zhou et al. and add new keywords specific to imperative and OO DL programming and *TensorFlow* 2. For instance, if we encounter a functor (either `__call__` or `call`, (e.g., Line 26, Listing 2)), we explore the class hierarchy to ensure that the class inherits from `tf.keras.Model`. Like Zhou et al., we inform developers of any contextual assumptions made during the analysis; developers can examine them during refactoring.

*3) Dataflow Analysis:* To track (imperative) tensor types, we adapt *Ariadne* [4], which operates on deferred-execution–style DL programs, to work with OO imperative DL code. *Ariadne* produces a dataflow graph as part of a pointer analysis and call graph construction. A dataflow graph summarizes the flow of objects and values in the program—abstracting possible program behavior—and is defined as follows [4]. Note that Python does not distinguish objects from values:

**Definition 1** (Dataflow Graph)**.** A dataflow graph $\mathcal{G} = \langle V, S, \prec \rangle$ where $V$ is the set of program variables, $S(v)$ is the set of objects possibly held by $v \in V$, and $x \prec y$ iff there is a potential dataflow from $y \in V$ to $x \in V$, e.g., via an assignment or function call.

Given a dataflow graph $\mathcal{G}$, we define a tensor *estimate* $T(v)$ as the set of possible tensor types held by $v$. The symbol $\mathcal{T}$ denotes the documented tensor type of the data source (q.v. Table III). This is implemented directly using the dataflow analysis in *Ariadne*, which we augmented for modern, imperative DL programs, and is defined as follows [4]:

**Definition 2** (Imperative Tensor Estimate)**.** Given a dataflow graph $\mathcal{G}$, a tensor *estimate* $\mathscr{T}(\mathcal{G}) = \langle T \rangle$ defines the set of tensor types a variable may take on. The type is defined as either the given data source type, dataflow in the program, or the result of other *TensorFlow* 2 APIs:

$$T(y) \quad \subseteq \quad \begin{cases} \{\mathcal{T}\} & y \text{ is a data source} \\ T(x) & y \prec x \\ \ldots & y \prec \text{ other } \textit{TensorFlow} \text{ 2 APIs} \end{cases}$$

Table III shows example tensor "generators" for imperative DL code that are the sources of the interprocedural dataflow analysis. A complete list may be found in our replication package [54]. A key difference here is that `Session` objects are no longer (exclusively) used for computation; thus, legacy API like `tf.placeholder()` are no longer useful. As the analysis is exclusively client-side, tensor creations are approximated by distinguishing APIs creating *new* tensors from those manipulating or creating new tensors *based on existing* tensors arguments. The former are considered tensor generators, while the latter are not. For example, `tf.constant()` creates a new tensor based on a literal or existing tensor argument, while `tf.add()` creates a new tensor based on two existing tensor arguments. Column **API** is the generator name, representing either a constructor or a value-returning function. Column **alias** is the "main" API alias, which we also consider a generator. Column **description** describes the API, and column **tensl** is *true* iff the API returns a "tensor-like" object, e.g., `tf.Variable`, as opposed to an actual tensor.

*a) Tracking Tensors in Containers:* Tensors may also reside in containers, which are generally difficult to track [61], [62]. We extend *Ariadne*'s dataflow analysis to also track tensors in (multidimensional) `tf.data.Datasets`—a popular API for processing tensor collections [24]. Table IV depicts example dataset generating API. Unlike Table III, these API may generate *datasets* from existing tensors.

```
1 def f():
2   return x ** 2 + y
3 x = tf.constant([-2, -3])
4 y = tf.Variable([3, -2])
5 f()
```

Listing 7: Lexical scoping [63].

*b) Dynamic Features:* Functions using dynamic Python features may have *implicit* `Tensor` parameters. For example, Listing 7 uses lexical scoping; because `f()` is called after its declaration, `x` and `y` are in the *closure* of `f()`. Thus, `x` and `y` at line 5 become *implicit* `Tensor` parameters of `f()`. We currently do not consider lexically-scoped tensor parameters; however, many real-world Python programs do not take advantage of many advanced dynamic features [5], [59]. Nevertheless, *Ariadne* supports several popular dynamic features—including those we have contributed (q.v. §V-A)—such as higher-order functions (callbacks), closures, decorators, and pointer analysis for field references (`x = obj.f`) and dictionary accesses (e.g., `x_dict['images']`) [4]. Moreover, *WALA* has been used on dynamic languages other than Python, e.g., for security analysis [64]. Other dynamic features such as introspection (e.g., `getattr()`) and code generation (e.g., `exec()`) are unsupported and are analogous to analyses of reflection in static languages that operate under a closed-world assumption [65]. *Ariadne* supports some metaprogramming, e.g., decorators; however, our subjects did not exhibit an abundance of other metaprogramming and dynamic computation graphs (q.v. §V), and we were able to refactor 326 (42.56% of) functions (q.v., §V-B4).

### D. Tracking Literal Function Arguments

To track potential literals arguments, we use a dataflow analysis from *Ariadne* to implement a constant propagation [66]. Literals are tracked through scalars, non-scalars, and objects of user-defined classes. The latter two are tracked through object fields; if literals flow to any field of an object, the object is considered to contain a literal value.

### E. Inferring Python Side-effects

We implement a novel side-effect static analysis for Python to infer functions containing Python side-effects (e.g., Listings 3 and 4). A function contains *Python* side-effects—those *not* caused by *TensorFlow* operations—iff Python operations cause heap locations, e.g., global variables, argument references, not allocated by the function to change as a result of its execution. The relationship is transitive; a function contains side-effects if any of its callees contain side-effects.

We conservatively approximate side-effects using a ModRef analysis that analyzes Python operations. ModRef analysis determines whether a function modifies (Mod) or references (Ref) a memory location [66]. We adapt the ModRef analysis of *Ariadne* to track Python side-effects. The analysis is interprocedural and context-sensitive, flow-insensitive, field-sensitive, and path-insensitive.

Heap locations are associated with call graph nodes where heap memory is allocated via call sites. If memory is allocated by a function and consequently altered by the *same* function via a Python expression or statement, the function does *not* have Python side-effects. Conversely, if a call site allocating heap memory resides *outside* the function's body and that location is modified, the function *has* side-effects. Memory allocated and consequently modified within a function's transitive closure is not considered a side-effect. Even if such memory "escapes" a function, we are only concerned with the function's execution for hybridization purposes. However, memory residing in global variables, argument references, or instance fields modified by the function *is* considered to be a side-effect. We also model the built-in `print()` function as side-effecting; such functions may alter global state (e.g., standard output). Moreover, we add built-in method summaries, e.g., `list.append()`, to *Ariadne* as mutating methods, affecting their receiver.

### F. Identifying Recursive Python Functions

To approximate recursive functions, we first augment *Ariadne*'s call graph construction algorithm to support popular idioms of imperative and OO DL code, namely, callable objects (functors) and related library callbacks. An example of the former is a class that implements the `__call__()` method, allowing instances of the class to be called as functions (e.g., Line 26, Listing 2). An example of the latter is a class that subclasses `tf.keras.Model` and overrides the `call()` method (e.g., Line 2, Listing 6). Such classes will inherit a `__call__()` method that invokes `call()` internally. For the former case, edges are added from call sites invoking instances of such classes to the `__call__()` method. A points-to analysis using the "function" name is performed to determine whether a call site is likely invoking an instance of a class or a regular function. For the latter case, edges are added from call sites invoking `__call__()` methods to `call()` methods. We then perform a depth-first search (DFS) on the call graph starting from the function node. To avoid infinite loops, a "seen" list of previously encountered nodes is maintained and continuously checked.

### G. Generalization Beyond TensorFlow

Due to its focus on speed of production models and extensive analysis [8], [9], [11], [23], [24], [48]–[53], we focus on hybridization in *TensorFlow*. However, other imperative DL frameworks, e.g., *MXNet* [67], *PyTorch* [13], have similar technologies, e.g., *Hybridize* [19], *TorchScript* [16]. For example, *PyTorch* has *TorchScript* that uses a similar decorator, `@torch.jit.script`, which compiles a Python function into a *TorchScript* graph [68]. To generalize our approach to other technologies, we would also consider functions decorated with such decorators as hybrid. We would then model *PyTorch* tensor operations adding a corresponding library summary to *Ariadne* as we have done in §IV-C3a for *TensorFlow*. For instance, Tables III and IV would be recreated for *PyTorch* APIs, e.g., using the *torch.utils.data* [69] documentation.

Because the underlying technology of `torch.jit.script` is slightly different than `tf.function`, another consideration is that there may be different rules that govern its usage. For example, it requires adherence to the *TorchScript* subset of Python, which might involve either stricter preconditions or additional code modifications for unsupported features. The former would necessitate augmenting our analysis with additional checks, while the latter would either involve incorporating more automated transformations or detailed failure statuses for manual inspection. We plan to extend our approach to support other frameworks in the future.

## V. EVALUATION

### A. Implementation

Our approach is implemented as a publicly available, open-source *PyDev* [1] *Eclipse* [2] IDE plug-in called HYBRIDIZE FUNCTIONS [27] that integrates the *WALA* [3] *Ariadne* [4] analysis framework (Fig. 5).

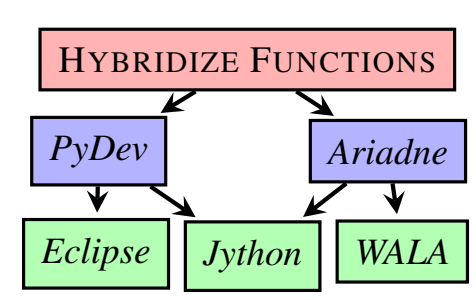

Fig. 5: Architecture.

*PyDev* provides efficient program entity indexing and extensive refactoring support, while *Eclipse* offers the refactoring framework and test engine [71]. We built atop *PyDev* a fully-qualified name (FQN) lookup feature for resolving decorator names. *WALA* provides static analyses, e.g., ModRef for side-effects (q.v. §IV-E), and *Ariadne* offers Python and tensor analysis. For transformation, *PyDev* ASTs with source symbol bindings serve as an intermediate representation, while static analysis uses SSA form IR [72]. Additional details are in our formal tool demonstration [27].

Both *PyDev* and *Ariadne* use *Jython* [70] for generating Python ASTs. While there is some redundancy, the ASTs serve

TABLE V: Experimental results.

| subject | dom | type | model | KLOC | fnc | rft | P1 | P2 | t (s) |
|---|---|---|---|---|---|---|---|---|---|
| chatbot | NLP | app | S2S | 0.82 | 16 | 14 | 14 | 0 | 4.13 |
| deep_recom | Recom. | lib | WDL | 3.07 | 30 | 16 | 16 | 0 | 53.93 |
| eth-nlu-nlm | NLP | app | RNN | 0.68 | 9 | 3 | 3 | 0 | 11.21 |
| GPflow | ML | lib | GP | 30.24 | 221 | 48 | 26 | 22 | 285.18 |
| gpt-2-tensorflow2 | NLP | app | GPT | 0.88 | 11 | 7 | 7 | 0 | 9.58 |
| lczero-training | Chess | app | α0 | 3.92 | 27 | 15 | 13 | 2 | 48.91 |
| mead-baseline | NLP | lib | Var. | 36.72 | 76 | 8 | 8 | 0 | 257.02 |
| MusicTrans-TF2 | Audio | app | MT | 1.74 | 11 | 7 | 7 | 0 | 13.61 |
| nlp-journey | NLP | app | Var. | 2.49 | 5 | 3 | 3 | 0 | 21.12 |
| NLPGNN | NLP | lib | GNN | 7.72 | 74 | 44 | 44 | 0 | 148.24 |
| nobrainer | CV | lib | CNN | 11.63 | 31 | 10 | 10 | 0 | 187.12 |
| ResNet50* | CV | lib | CNN | 7.60 | 43 | 15 | 15 | 0 | 157.63 |
| samples | Var. | ref | Var. | 3.98 | 17 | 6 | 6 | 0 | 57.00 |
| TF-Examples | Var. | ref | Var. | 1.79 | 53 | 48 | 47 | 1 | 58.80 |
| tf-yolov4-tflite | CV | tk | CNN | 2.74 | 9 | 7 | 7 | 0 | 30.34 |
| TF2.0-Examples | CV | ref | Var. | 3.45 | 36 | 21 | 21 | 0 | 55.54 |
| TensorflowASR | NLP | tk | CNN | 10.31 | 67 | 25 | 25 | 0 | 100.31 |
| tf-dropblock | ML | lib | CNN | 0.12 | 2 | 2 | 2 | 0 | 0.75 |
| tf-eager-fasterrcnn | CV | tk | CNN | 2.16 | 28 | 27 | 27 | 0 | 65.62 |
| Total | | | | 132 | 766 | 326 | 301 | 25 | 1,566 |

* ResNet50 is from *TensorFlow Model Garden* [73].

TABLE VI: Refactoring failures.

| | failure | pc | cnt |
|---|---|---|---|
| F1 | No primitive parameter | P3 | 51 |
| F2 | Primitive parameter(s) | P1 | 59 |
| F3 | Side-effects | P1/P2/P3 | 82 |
| F4 | Missing CG node | P1/P2/P3 | 272 |
| | Total | | 464 |

different purposes (transformation vs. analysis). We match *PyDev* ASTs with *Ariadne* SSA IR by matching filenames, functions, line numbers, and parameter positions, employing a technique similar to Khatchadourian et al. [31] for Java software. Some representation differences between the tools complicate matching, e.g., *Ariadne* considers type hints part of parameter expressions while *PyDev* does not.

We augment *Ariadne* by vastly expanding the XML library summaries to support popular imperative DL APIs and adding support for various Python language features widely used in imperative DL programs, including module packages, wild card imports, intra-package references, package initialization scripts, automated unit test entry points discovery, iteration of non-scalar tensor datasets, additional library modeling, and analysis of static and class methods, custom decorators, and callable objects. We contribute these back to *Ariadne*.

Since there are only two possible values with minimal retracing risk, we optionally consider Booleans during literal inference. Other options include whether to consider *pytest* entry points and to always follow type hints regardless of hybridization arguments during tensor inference.

## B. Experimental Evaluation

To validate our approach, we conduct an experimental evaluation that addresses the following research questions:

**RQ1.** ***Effectiveness*** How many candidate functions can the approach successfully refactor for hybridization, and what is the distribution of precondition failures?

**RQ2.** ***Performance Impact*** What performance improvements (if any) are achieved by the automated hybridization refactorings?

**RQ3.** ***Correctness*** Do the refactorings preserve program semantics, particularly in terms of model accuracy?

**RQ4.** ***Scalability*** What is the computational cost of the analysis, and how does it scale with code size?

*1) Subject Selection:* Our evaluation involves nineteen open-source Python imperative DL systems (Table V) of varying size and domain and include a mix of ML libraries and frameworks, toolkits, reference implementations, applications, and model architectures. Since we focus on hybridization in *TensorFlow*, we choose *TensorFlow* (v2)-based projects. We aim to study diverse DL systems with available optimization opportunities. Such systems may have been originally developed imperatively but largely execute eagerly or suboptimally either because developers do not know of hybridization or do not know how to use it correctly [23]–[25]. They may also be migrated legacy (TF1) systems and now need to recover run-time performance. Subjects must have at least one function that: (i) has at least one tensor or tensor-like parameter or (ii) is hybrid, i.e., a candidate function. Column **KLOC** denotes the thousands of source lines of code, ranging from 0.12 to 36.72. Subjects with unit tests had identical test results before and after refactoring. GitHub metrics may be found in our dataset [54]. While our library subjects are generally actively maintained, most open-source ML projects are not [74].

*2) Study Configuration:* The analysis is executed on an Intel Xeon i9 machine with 24 cores and 64 GB RAM. Column **t** is the running time in seconds, averaging 11.86 s/KLOC. This addresses **RQ4** (scalability), showing that our analysis scales reasonably with code size, requiring an average of 0.17 s per candidate function analyzed. We set options to: (i) discover *pytest* entry points, (ii) not consider Booleans during literal inference, (iii) use speculative analysis, and (iv) always follow type hints. Some subjects (e.g., [60]) only include notebooks, for which we first convert to Python files using *ipynb-py-convert*. Minor manual editing was sometimes required to complete the conversion; directly processing notebooks is for future work. Using *2to3*, minor edits were made in some cases to upgrade code to use Python 3. Some subjects, e.g., *tf-dropblock*, were libraries that include driver code in `README.md` files; "fenced" code was copied into (main) Python (driver) files. Some subjects (e.g., [76], [77]) were missing (empty) package initialization scripts, which were added manually.

*3) Refactoring Results:* To address **RQ1** (effectiveness), we analyze 766 Python functions (column **fnc**, Table V) across nineteen subjects. Of those, we automatically refactored 42.56% (column **rft**) despite being highly conservative. These functions have passed all preconditions; those not passing preconditions are not transformed (cf. Table VI). Columns **P1–2** are functions passing corresponding preconditions (cf. Tables I and II). Column **P3** is omitted as all values are 0.

> ***Answer to RQ1***: Our approach successfully refactors 42.56% of candidate functions across diverse DL projects, demonstrating reasonable effectiveness despite a conservative analysis.

*4) Refactoring Failures:* Continuing our analysis of **RQ1** (effectiveness), Table VI categorizes reasons why functions

cannot be refactored (column **failure**), potential corresponding preconditions (column **pc**), and respective counts (column **cnt**). Note that a function may be associated with multiple failures. There were 464 failures across all subjects. Side-effects (F3, Listings 3 and 4), potentially affecting every precondition, accounted for 17.67%.

The analysis provides detailed insight into the distribution of precondition failures requested in **RQ1**, with side-effects being the most common barrier to refactoring. *Having* primitive parameters (F2, 12.72%) prevents a function from being hybridized (P1), e.g., `train()` in Listing 5a would be included in F2 due to `num_steps` had it not originally been hybrid. F1, at 10.99%, is the least common and is due to having *no* primitive parameters—unlike F2, F1 prevents a function from being *de-hybridized*. Function `train()` in Listing 5a would be an example of F1 had it not had `num_steps`. Missing call graph (CG) nodes (F4, 58.62%), also potentially affecting each precondition, arise when functions are: (i) unreachable from entry points, (ii) in libraries or frameworks, or (iii) called either by unsupported language features or dynamically, e.g., using `getattr()`. Note that function callbacks are not necessarily problematic as *Ariadne* resolves them with the exception of missing external library modeling. Nevertheless, we still can refactor 326 (42.56% of) functions despite being conservative.

*5) Refactoring Warnings:* Hybrid functions that potentially contain side-effects are not transformed per Table II to preserve semantics. However, such functions may be buggy (q.v. §IV-A2), and we issue refactoring "warnings" for these. During our evaluation, we discovered fifteen hybrid functions with potential side-effects across three subjects. While we focus on improving nonfunctional aspects, automated bug detection is potential future work (q.v. §VII).

*6) Performance Evaluation:* Many factors can influence run-time performance. Nevertheless, to address **RQ2** (performance impact) and **RQ3** (correctness), we assess the run-time performance impact and semantic preservation of our refactorings. Similar manual refactorings (Listing 2) reduce training time by ∼89% on modest datasets (§III).

*a) Benchmarks:* To assess run-time performance, we use models from a subset of subjects in Table V. Unfortunately, none include dedicated performance tests; we thus focus on model training time, as it tends to dominate the DL pipeline. We add timing metrics and average model *accuracy* and *loss* per epoch to each benchmark where applicable. *Lost accuracy* is the difference between original and refactored model accuracies. Our transformations are not intended to improve model accuracy but rather speed. Using model accuracy to assess DL code refactoring is standard practice [78]–[81].

Subjects were chosen such that: (i) code did not require significant manual intervention and ran to completion, (ii) library or framework code had available tests or examples, (iii) sample datasets were either provided or comparable alternatives were locatable, (iv) code included minimal calls to deprecated API calls, (v) the benchmark file trains a DNN, and (vi) the benchmark file includes refactorings performed by our tool. We sometimes increased epochs to better resemble real-world workloads—a common practice [28], [31], [82]—as insufficient execution repetitions can negatively impact performance assessments [83]. Other times, we decreased epochs for tractability when the same speedup held beyond a threshold. We closely followed methodologies of previous work and original developer guidelines while keeping training tractable. To minimize bias, the epochs used were the same for both original and refactored versions. We expect the performance ratio to hold as epochs increase.

TABLE VII: Average run times of DL benchmarks.

| # benchmark | Keps | oa | ra | ol | rl | ot | rt | su |
|---|---|---|---|---|---|---|---|---|
| 1 test_dcn | | | | | | 0.58 | 0.56 | 1.03 |
| 2 test_transformer | | | | | | 5.88 | 3.83 | 1.53 |
| 3 train_deepfm | 0.01 | | | 0.56 | 0.56 | 117 | 110 | 1.06 |
| 4 train_trans | 0.01 | 80.18 | 81.70 | 0.57 | 0.56 | 87.22 | 86.86 | 1.00 |
| 5 gpt2_model | 0.1 | 16.42 | 16.13 | 6.74 | 6.73 | 90.69 | 71.37 | 1.27 |
| 6 train | 0.01 | 2.33 | 2.38 | 4.96 | 4.92 | 1,330 | 919 | 1.45 |
| 7 GAAE | 0.1 | 69.03 | 69.22 | 66,270 | 66,241 | 53.44 | 45.54 | 1.17 |
| 8 train_gan | 1 | 79.26 | 78.80 | 1.18 | 1.18 | 35.05 | 19.53 | 1.79 |
| 9 autoencoder | 20 | | | 0.01 | 0.01 | 111 | 34.25 | 3.25 |
| 10 bidirectional_rnn | 1 | 82.09 | 81.75 | 0.58 | 0.59 | 27.89 | 4.79 | 5.82 |
| 11 custom_layers | 5 | 94.16 | 93.99 | 3.28 | 3.28 | 12.53 | 5.30 | 2.37 |
| 12 convo_net | 2 | 98.68 | 98.70 | 1.48 | 1.48 | 31.08 | 17.71 | 1.75 |
| 13 dcgan | 0.5 | | | 1.22 | 1.20 | 77.97 | 59.84 | 1.30 |
| 14 dynamic_rnn | 2 | 85.80 | 86.58 | 0.30 | 0.29 | 48.15 | 8.49 | 5.67 |
| 15 logistic_regress | 10 | 88.65 | 88.61 | 0.46 | 0.46 | 11.44 | 3.81 | 3.01 |
| 16 multigpu_train | 1 | | | 1.67 | | 9,285 | | |
| 17 neural_network | 20 | 99.33 | 99.33 | 0.03 | 0.03 | 48.81 | 24.54 | 1.99 |
| 18 recurrent_net | 3 | 87.27 | 87.29 | 0.42 | 0.41 | 42.69 | 7.52 | 5.68 |
| 19 save_res_model | 10 | 94.04 | 94.16 | 51.94 | 51.66 | 40.59 | 14.26 | 2.85 |
| 20 tensorboard_ex | 3 | 87.27 | 87.12 | 110.24 | 112.08 | 8.79 | 4.57 | 1.92 |
| 21 autoencoder | | 98.62 | 98.60 | 0.10 | 0.10 | 11.40 | 10.14 | 1.12 |
| 22 CNN | 0.01 | 84.94 | 84.38 | 0.04 | 0.05 | 32.85 | 32.98 | 1.00 |
| 23 Multilayer | | 78.63 | 77.70 | 54.02 | 54.33 | 8.28 | 3.94 | 2.10 |
| 24 ResNet18 | 0.01 | | | 0.74 | 0.76 | 34.35 | 34.65 | 0.99 |
| 25 RPN/train | 0.01 | | | 0.09 | 0.09 | 1,359 | 1,043 | 1.30 |
| 26 YOLOV3/train | 0.01 | | | 329.97 | 325.02 | 920 | 572 | 1.61 |

Minor manual changes were made to some subjects to run them with recent *TensorFlow* versions. Such APIs were either deprecated or replaced with more native equivalent *TensorFlow* APIs not previously available in earlier versions (i.e., < 2.9.3). For such changes, we submitted pull requests, some of which were merged. These changes were needed to run the subjects but not necessarily to analyze them as our approach is capable of handling multiple DL library versions through library modeling. Several manual transformations were made to avert *TensorFlow* bugs, including pending bugs [84], numerical instability [85], and one (pending) crash [86], [87] related to Adam optimizers [88] and software layering within the *TensorFlow* [23]. We also removed early stopping [89] on two benchmarks so that the original and refactored versions could be compared fairly. For one benchmark, we manually added `reduce_retracing=True` [63] to a new `@tf.function` decorator after *TensorFlow* (early on) reported retracing due to varying tensor argument dimensions. In the future, we will explore automatically adding decorator arguments.

*b) Results:* Table VII reports the average run times of five runs in seconds. Benchmarks 1 to 4 are for *deep_recommenders* [90], benchmark 5 is for *gpt-2-tensorflow2.0* [91], benchmark 6 is for *MusicTransformer* [92], benchmarks 7 to 8 are for *NLPGNN* [76], benchmarks 9 to 20 are for *TensorFlow-Examples* [60], and benchmarks 21 to 26 are for *TensorFlow2.0-Examples* [93]; benchmark names have been shortened for brevity. Columns **oa** and **ra** are the average original and refactored model accuracies per epoch, respectively, in percentages. Columns **ol** and **rl** are the total original and refactored model losses per epoch, respectively. Some benchmarks measured different kinds of model losses, which we averaged—a common practice [94]–[96]. Columns **ot** and **rt** are the original and refactored run times in seconds, respectively, and column **su** is the relative speedup ($runtime_{old}/runtime_{new}$). The resulting average relative speedup from our refactoring is 2.16x.

> ***Answer to RQ2***: Our refactorings achieve an average relative speedup of 2.16x across performance benchmarks, demonstrating measurable performance improvements.

> ***Answer to RQ3***: The refactorings result in negligible accuracy loss (0.03%) and gained loss (–0.05%).

*7) Discussion:* The results demonstrate practical feasibility: our tool refactored 42.56% of candidate functions, achieving 2.16x average relative speedup with negligible accuracy loss (0.03%) and gained loss (–0.05%). This suggests that any side-effects of the refactoring are minimal if at all, making it suitable for real-world use.

> ***RQ Answer Summary***: **RQ1** – Our approach successfully refactored 42.56% of functions with side-effects being the primary barrier. **RQ2** – We achieved 2.16x average speedup. **RQ3** – Semantic preservation was confirmed with negligible accuracy differences. **RQ4** – The analysis scales reasonably, averaging 11.86 s/KLOC.

*a) Refactoring Failures:* From Table VI, F3 accounted for 17.67% of failures, indicating that side-effects are common in practice and motivating future work to help developers avoid them. F2 (12.72%) involves primitive parameters requiring significant redesign, while F1 (10.99%) prevents de-hybridization of already-optimal functions. F4 (58.62%) may be due to dead code or limitations addressable through improved library modeling and dynamic analysis integration.

*b) Performance Evaluation:* The average relative speedup of 3.24x obtained from *TensorFlow-Examples* reflects that it contains the most refactorings (48/53; 90.57%), while *deep_recommenders* achieved 1.16x with fewer refactorings (16/30; 53.33%). This suggests that more refactored functions lead to greater performance improvements [97]. Some benchmarks had speedups at or just under 1 due to existing `tf.function` usage, slight overhead, or unsoundness, while others achieved significant speedups by identifying additional optimization opportunities.

*c) Unsoundness:* Unsoundness may cause missed optimization opportunities, resulting in relative speedup closer to 1. We employ a conservative approach when analyzing this dynamic language. We observed no performance degradation and negligible differences in model accuracy and loss, likely due to conservativeness and that many real-world Python programs do not extensively use advanced dynamic features [5], [59]. Our tool provides analysis assumptions as refactoring "info" statuses for developer review.

*d) Complex Models:* Under certain conditions, arguments to the `__call__()` method of *Keras* models are automatically cast to tensors. For example, in benchmark 13, numpy arrays are sent to `Generator.call()`; *Keras* then casts the numpy array to a tensor before executing the method. In the calling context, it is an numpy array, but in the function definition, it is a tensor. The controlling `autocast` flag in this case is obtained from an environmental variable. Our analysis does not track this tensor; consequently, the corresponding method is determined not to have a tensor parameter despite the model potentially benefiting from hybridization. Nevertheless, benchmark 13 still achieved a relative speedup of 1.30x as other constituent models were hybridized by our tool.

*e) Framework-Invoked Hybridization:* Some *Keras* APIs, e.g., `Model.fit()`, call `tf.function` internally. For example, in *deep_recommenders* [90], some benchmarks did not have a considerable speedup; however, in subclassed *Keras* models, `fit()` may not *always* be called. The model's `call()` method, for instance, may be invoked instead, which is *not* automatically hybridized. Our tool may also hybridize *other* functions that are *not* invoked by `fit()`.

### C. Threats to Validity

*a) Internal Validity:* Static analysis has limitations regarding Python dynamic features (q.v. §IV-C3b); however, we use a best-effort, conservative approach that fails refactoring preconditions and halts transformations if the analysis is inconclusive. During the performance evaluation (§V-B6), we observed negligible differences in model accuracy and loss. Speculative analysis (§IV-C2) has previously been found to be reliable and includes any assumptions made that developers can examine [5].

Our approach may incorrectly classify functions due to analysis imprecision. To mitigate this, we adopt conservative heuristics that prefer false negatives (missed opportunities) over false positives (incorrect transformations). We also validate transformations by checking that model accuracy is preserved and performance is improved.

*b) External Validity:* Subjects may not be representative of imperative DL systems. To mitigate this, subjects were chosen from diverse domains and sizes, have been used in previous studies [8], [9], [11], [23], [28], [48], [49], [51], [52], [98], [99], and are included in data science-specific datasets [100]. Subjects also include lesser-known repositories to understand hybridization opportunities available to the DL community-at-large. While some subjects include sample code, they serve as reference implementations with non-trivial GitHub metrics. Subjects included only *TensorFlow* clients; it is possible that clients of other imperative DL frameworks [13], [67] would yield different results. However, *TensorFlow* has, since inception, focused on speed of production models, has been extensively studied (q.v. §IV-G), and `tf.function` is popular [23]. *PyTorch* systems, which have historically executed eagerly, may not be as speed-critical. Subjects also do not include Python Notebooks; directly processing notebooks is for future work, and our experimentation involved first converting some notebooks to Python files.

*c) Construct Validity:* Our performance evaluation is based on execution time and model accuracy metrics, which

are standard measures in the DL performance literature. However, other factors such as memory usage, compilation time, and developer productivity are not explicitly measured. The generalizability of our precondition set may be limited to the specific patterns observed in our subject programs, though we drew from comprehensive *TensorFlow* documentation [22] to design them.

## VI. Related Work

*a) Performance Optimization & Analysis of DL Systems:* Cao et al. [24] characterize general performance bugs in DL systems. They find that developers often struggle with knowing where to add `@tf.function` and how to implement decorated functions for optimal performance. Their static checker detects some general performance problems; however, it does not consider hybridization issues. Gao et al. [101] study low GPU utilization of DL jobs. Beyond performance bugs, Castro Vélez et al. [23] detail challenges in migrating imperative DL code to graph execution.

Zhou et al. [5] use speculative analysis for optimizing the performance of *procedural*, deferred-execution–based analytics programs; we use it for optimizing *imperative* DL code via hybridization. Likewise, Islam [102] detects misuse using procedural-style call patterns.

*b) Automated Refactoring & API Migration:* Ni et al. [103] generate mappings between (different versions of) data science APIs for later use by automated refactorings. However, they either switch between APIs or migrate them to a new version. In contrast, we enhance non-functional facets of DL systems by improving the usage of constructs found in a particular API version.

Baker et al. [25] extract common *TensorFlow* API misuse patterns, one of which (and corresponding fix suggestion) involves (a specific use case of) `tf.function`. They do not, however, refactor eager functions to hybrid and vice-versa. Likewise, Wei et al. [53] investigate the characteristics of *TensorFlow* and *PyTorch* DL API misuse patterns but do not consider hybridization. Nikanjam and Khomh [50] catalog various design smells in DL systems and recommend suitable refactorings. Dilhara et al. [98] study ML library evolution and its resulting client-code modifications. Dilhara et al. [34] and Tang et al. [104] analyze repetitive code changes and refactorings made in ML systems, respectively. While valuable, these studies do not deal with automatically migrating imperative DL code to graph execution. Dilhara et al. [105] automate frequent code changes in general Python ML systems. To the best of our knowledge, their work does not consider side-effects, recursion, and other necessary analyses to increase the likelihood that hybridization is safe and potentially advantageous.

*c) Static Analysis of Python & Tensor Analysis:* Dolby et al. [4] build a static analysis framework for tensors shapes in procedural DL code, as do Lagouvardos et al. [41]. The latter approach is built from the former, as is ours. On the other hand, Zheng and Sen [46] use dynamic analysis to infer likely tensor shapes, whereas we use static analysis to infer tensor parameters, side-effects, and others. Their work is complementary and could be used to improve our tensor parameter inference (q.v. §VII).

Mukherjee et al. [32] perform static analysis of arbitrary Python code that use AWS APIs. Like us, they also use speculative analysis [5] as a fallback to resolve Python's dynamic features. Rak-amnouykit et al. [106] develop a hybrid Andersen-style points-to analysis for Python.

*d) General Refactoring Research:* Side-effect analysis is found in other contexts, such as refactoring to parallelize sequential code [30], [31] and in other dynamic languages [107]. Other refactorings enhance program structure [108], upgrade to new API versions and design patterns [109]–[112], improve energy consumption [113], eliminate code redundancy [114], make mobile applications more asynchronous [115], migrate to cloud-based microservices [116], and others [117], [118].

## VII. Conclusion & Future Work

Our automated refactoring approach assists developers in determining which otherwise eagerly-executed imperative DL functions could be effectively and efficiently executed as graphs. The approach features novel static imperative tensor and side-effect analyses for Python. A speculative (keyword-based) analysis is used to resolve difficult cases caused by Python's dynamism that informs developers of any assumptions made. The approach was implemented as a *PyDev Eclipse* IDE plug-in and evaluated on nineteen open-source programs, where 326 of 766 candidate Python functions (42.56%) were refactored. A performance analysis indicated an average relative speedup of 2.16x.

In the future, we will explore repairing hybridization bugs, potentially modifying decorator arguments by augmenting *Ariadne* to infer tensor shapes in imperative DL programs, integrating dynamic analyses [45], [46], directly processing Python notebooks using *LSP* [119], and supporting explicit calls to `tf.function()`. We will also explore extending our approach to other frameworks via NLP of API documentation to generate library summaries or via LLMs.

Per §IV-A, precondition formulation is hinged on the usage constraints of hybridization. While our tool was able to refactor 42.56% of candidate functions (Table V), it may be possible to transform more candidates using more sophisticated analyses. For example, F3 (Table VI) could be reduced by rewriting functions to be pure as discussed in §V-B7a. In the future, we will explore this possibility by first examining some simpler cases (e.g., replacing `print()` with `tf.print()`). We also plan to extend our evaluation by comparing our automated approach to manual refactorings. One potential challenge here, however, is obtaining a reliable ground truth as developers have historically struggled with (manually) using hybridization correctly [17], [22]–[25]. Thus, we plan to perform a user study and compare the relative speedups, accuracy, and model losses of the two classes.

## Acknowledgment

We thank the anonymous reviewers and Medha Belwadi and Pranavi Gollanapalli for their work on data collection.